\documentclass[11pt,twoside]{article}
\usepackage{asp2010}

\usepackage{graphicx}

\resetcounters

\begin{document}

\title{LOFT (Large Observatory For X-ray Timing): a candidate X-ray mission for the next decade}
\author{M. Hernanz, on behalf of the LOFT collaboration \\ 
(see http://www.isdc.unige.ch/loft/index.php/loft-team/community-members)
\affil{Institute of Space Sciences - ICE (CSIC-IEEC), Campus UAB, Fac. Ci\`encies, C5 par 2$^a$ pl.,
08193 Bellaterra (Barcelona), Spain}
}

\begin{abstract}
LOFT is one of the four medium mission candidates (M3), selected by ESA in the framework of the Cosmic Vision 
Programme (2015-2025), for feasibility study. If approved by ESA in 2014, its launch is foreseen in 2022-2024. 
LOFT is being designed to observe X-ray sources with excellent temporal resolution and very good spectral 
capability. Its main objectives are to directly probe the motion of matter in the very close vicinity of black holes 
(Strong Field Gravity), as well as to study the physics of ultra dense matter (Neutron Stars). The payload includes 
a {\bf Large Area Detector (LAD)} and a {\bf Wide Field Monitor (WFM)}. The LAD is a collimated ($< 1$ degree field of view) 
experiment operating in the energy range 2-30 keV, with a 10 m$^2$ peak effective area and an energy resolution 
of 260 eV at 6 keV. The WFM will operate in almost the same energy range than the LAD, 2-50 keV, enabling 
simultaneous monitoring of a few-steradian wide field of view, with an angular resolution of  $< 5$ arcmin. In addition 
to its main scientific objectives, LOFT will also do a complete plan of observatory science, studying with 
unprecedented detail in the 2-80 keV range several transient phenomena, like accreting white dwarfs in cataclysmic 
variables, novae in outburst (internal and external shocks in the ejecta in classical novae, and shocks with the wind 
of the companion in symbiotic recurrent novae) and post-outburst novae (once accretion is re established). 
\end{abstract}

\section*{LOFT description}
\begin{figure}
\centerline{
\includegraphics[width=15cm]{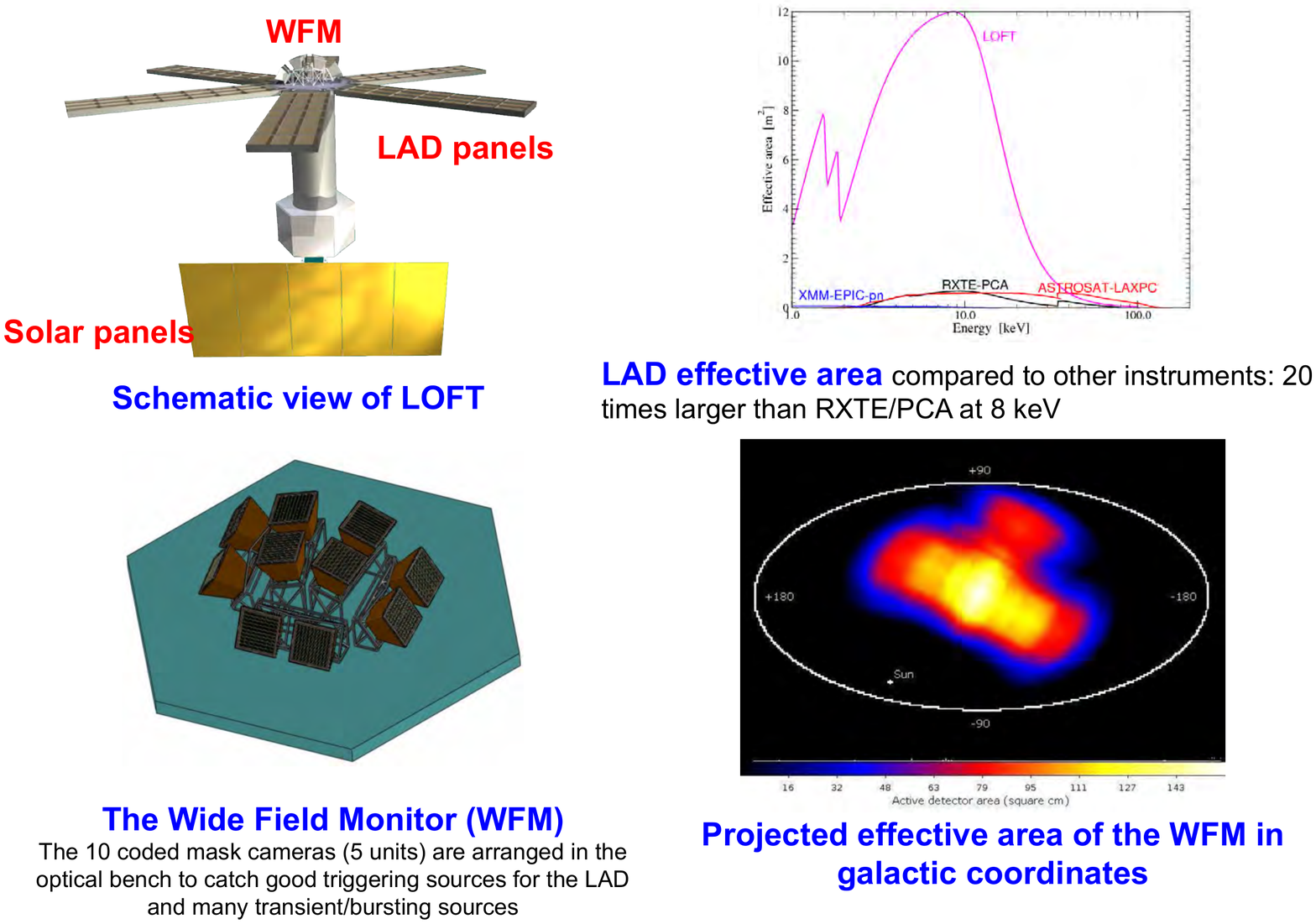}
}
\vspace{-1cm}
\caption{LOFT satellite and the WFM (left panels). Effective area of the LAD and of the WFM (right panels) }
\label{fig1}
\end{figure}

\begin{figure}
\centerline{
\includegraphics[width=15cm]{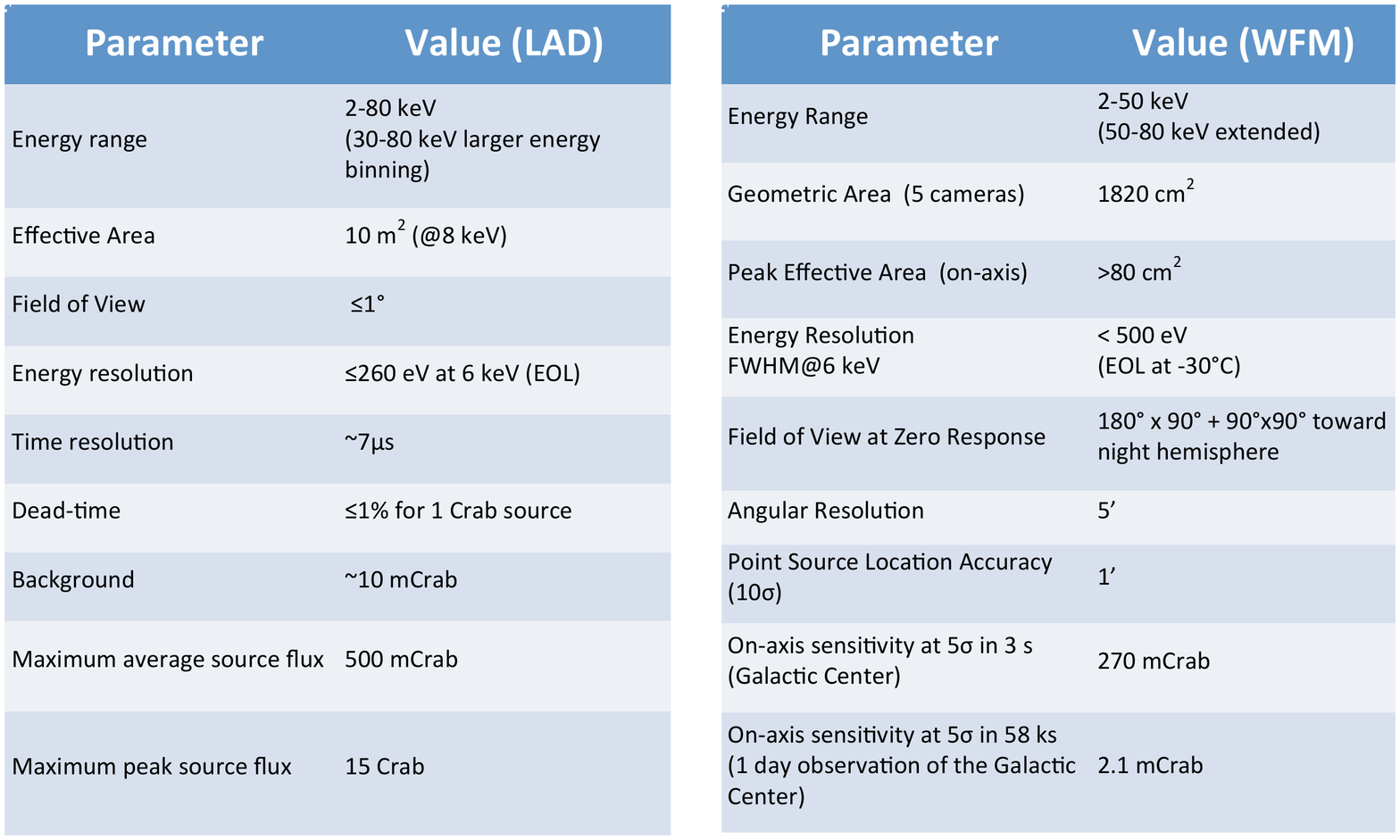}
}
\vspace{-2cm}
\caption{Performance of the LOFT instruments: LAD and WFM}
\label{fig2}
\end{figure}

In Figure \ref{fig1} we show a scheme of the LOFT satellite, with the LAD and WFM instruments, together with a detailed view of the 
configuration of the WFM, with its 10 coded mask based cameras. The effective area of the LAD instrument, as compared with that of 
previous similar instruments flown is also shown (an improvement by a factor of 20 at E=10 keV is expected), as well as the WFM field 
of view - depicted as the projected effective area in galactic coordinates. The performances of the two instruments for LOFT, LAD and WFM, 
are shown in Figure \ref{fig2}. 

\begin{figure}
\centerline{
\includegraphics[width=14cm]{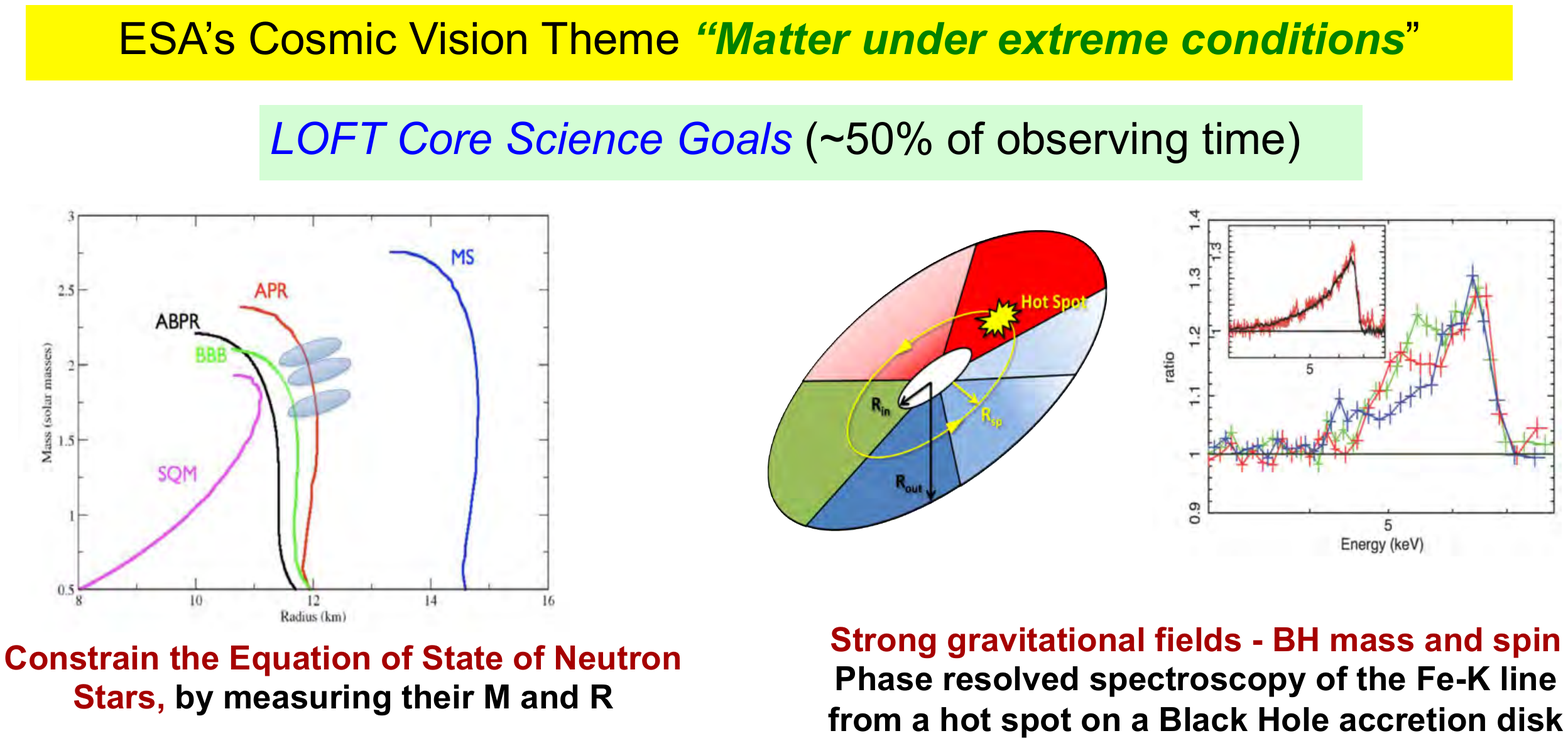}
}
\vspace{-3cm}
\caption{Main themes of the LOFT core science}
\label{fig3}
\end{figure}

\begin{figure}
\centerline{
\includegraphics[width=14cm]{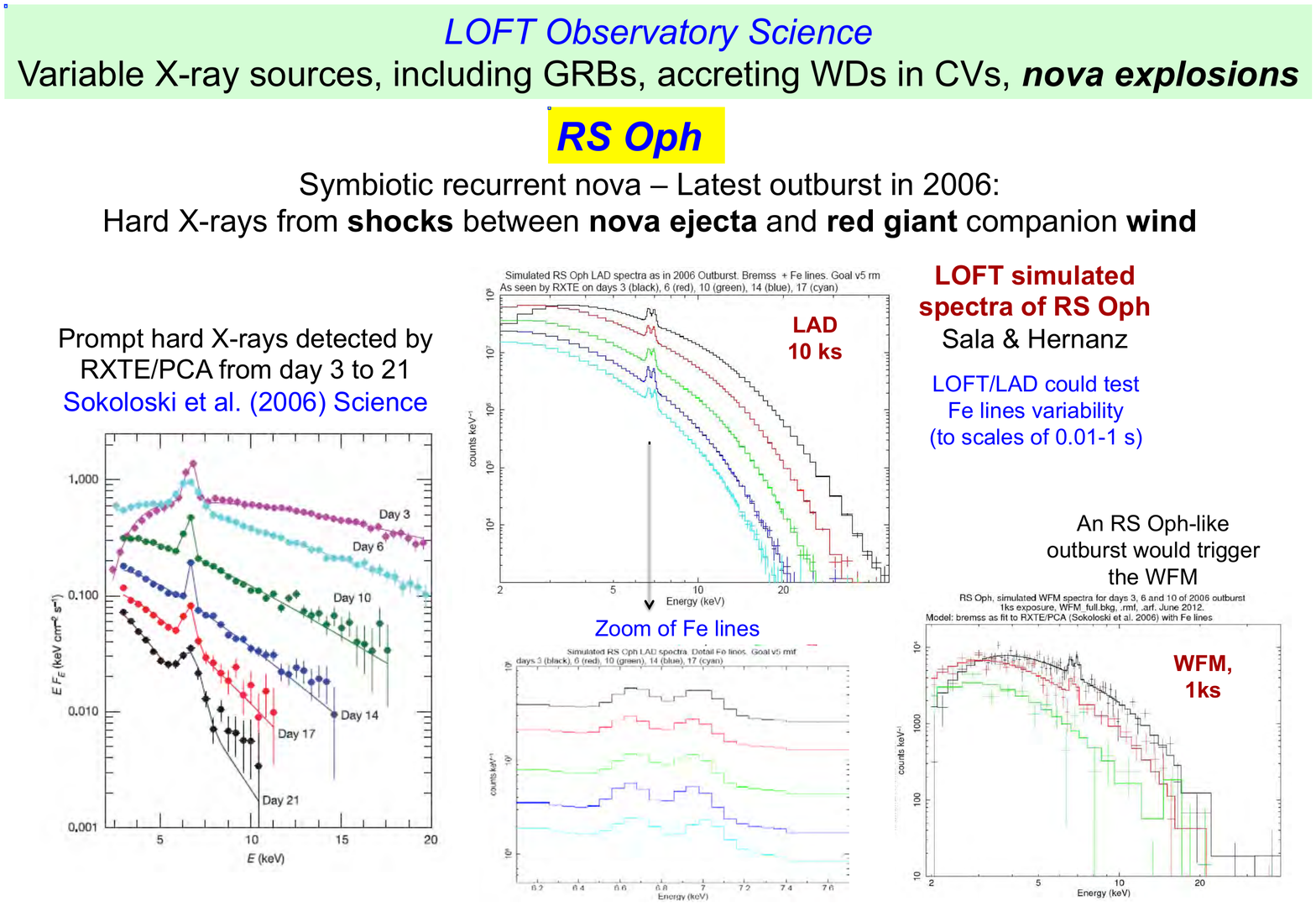}
}
\vspace{-1cm}
\caption{An example of LOFT observatory science: recurrent novae}
\label{fig4}
\end{figure}

The main scientific topics of LOFT are related to the theme ``Matter under extreme conditions", of the ESA Cosmic Vision Programe 2015-2025\\
\centerline{(see Figure \ref{fig3} and http://www.isdc.unige.ch/loft/ for details).}
The equation of state of Neutron Stars and the behavior of matter under strong field gravity (i.e., mainly matter accreting onto Black Holes) are the main 
scientific topics. 
In addition, LOFT will do what we call ``Observatory Science" (see Figure \ref{fig4}), one topic being the observation of the hard X-ray emission from classical and recurrent novae, related to the shocks between the ejecta and the circumstellar matter. This topic is especially relevant to understand the 
recently discovered ``Fermi novae", emitting in the GeV range. RS Oph (2006 eruption) would have been detected by the Fermi/LAT instrument; the information 
obtained by RXTE in 2006 was crucial to make this prediction. With LOFT, such a study would have been much more precise (see Figure \ref{fig4} and paper by Hernanz in this same volume).
  
A list of selected papers related to LOFT is presented in the references section.

\acknowledgements The author thanks funding from the Spanish MINECO project AYA2011-24704, FEDER funds and the AGAUR (Generalitat of Catalonia) 
project 2009 SGR 315.

\bibliography{aspauthor}

\end{document}